\documentclass[prd,twocolumn,
nofootinbib,preprintnumbers]{revtex4}

\usepackage{times}
\usepackage{amssymb,amsbsy,amsmath,amsfonts}
\usepackage{mathrsfs}
\usepackage{graphicx}
\usepackage{graphics,psfrag,longtable}
\usepackage{nicefrac}
\usepackage{esint}
\usepackage{upgreek}

\usepackage[colorlinks,linktoc=all,citecolor=blue,linkcolor=red]{hyperref} 

\DeclareMathAlphabet{\mathantt}{OML}{antt}{l}{it}
\DeclareMathAlphabet{\mathpzc}{OT1}{pzc}{m}{n}

\def\beq{\begin{equation}}
\def\eeq{\end{equation}}
\def\bea{\begin{eqnarray}}
\def\eea{\end{eqnarray}}
\def\beqa{\begin{equation}\begin{array}{l}}
\def\eeqa{\end{array}\end{equation}}
\def\eqlab#1{\label{eq:#1}}

\def\Eqref#1{Eq.~(\ref{eq:#1})}

\def\Figref#1{Fig.~\ref{fig:#1}}

\def\sixth{\mbox{$\frac{1}{6}$}}

\def\barr{\left(\begin{array}{c}}
\def\earr{\end{array}\right)}
\def\bmat{\left(\begin{array}{cc}}
\def\emat{\end{array}\right)}

  \def\eps{\epsilon}

\def\nn{\nonumber}
\def\dd{\mathrm{d}}

\def\3d{3-D}

\def\ol#1{\overline{#1}}

\begin{document}
\title {Reply to ``Comment on `Breakdown of the expansion of finite-size corrections to the hydrogen Lamb shift in moments of charge distribution' ''}

\author{Franziska Hagelstein}
\author{Vladimir Pascalutsa}
\affiliation{
Institut f\"ur Kernphysik, Cluster of Excellence PRISMA,  Johannes Gutenberg-Universit\"at Mainz, D-55128 Mainz, Germany}
\begin{abstract}
To comply with the critique of the Comment [J.~Arrington,
arXiv:1602.01461], we consider another
modification of the proton electric form factor, 
which resolves the ``proton-radius puzzle''. The proposed
modification satisfies all the consistency criteria put forward in the Comment,
and yet has a similar impact on the puzzle as that of the original paper.
Contrary to the concluding statement of the Comment,
it is not difficult to find an ad hoc modification
of the form factor at low $Q$ that resolves the discrepancy and
is consistent with analyticity constraints. We emphasize once again that
we do not consider such an ad hoc modification of the proton form factor
to be a solution of the puzzle until a physical mechanism for it is found.
\end{abstract}
\date{\today}
\maketitle
\begin{figure}[tbh] 
    \centering 
       \includegraphics[scale=0.48]{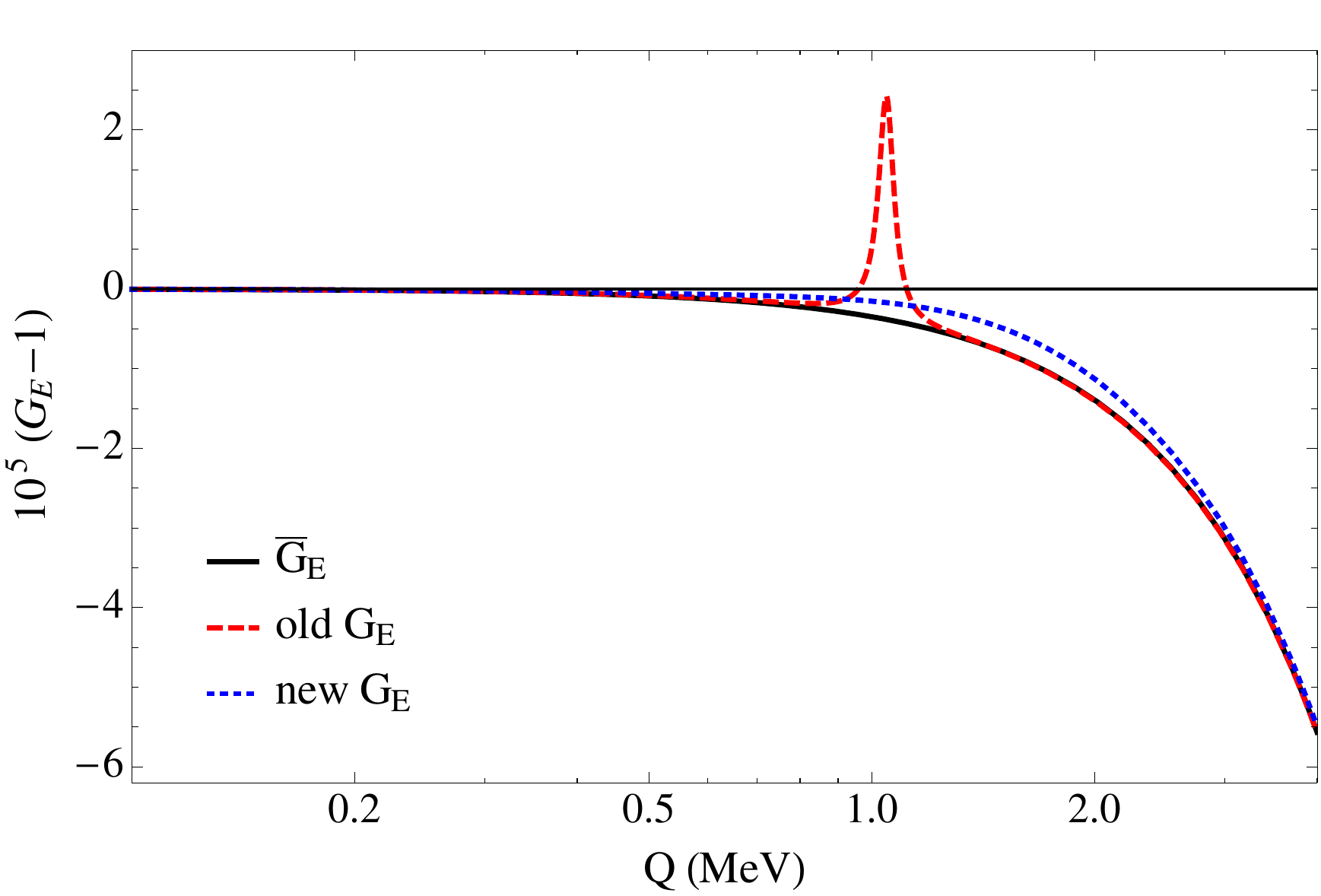}
       \caption{$G_E(Q^2)-1$ and $\ol G_E(Q^2)-1$ as a function of $Q$. The solid black curve shows the empirical FF from Ref.~\cite{{Arrington:2006hm}}. The dashed red curve  shows the modified FF from Ref.~\cite{Hagelstein:2015yma}. The dotted blue curve is the modified FF of this work.}
              \label{fig:GEminus1}
\end{figure}

The formalism developed in Ref.~\cite{Hagelstein:2015yma} was
illustrated by a modification of the proton electric form factor (FF),
$G_E$, which could reconcile the discrepancy in the various proton
radius extractions. 
As is correctly pointed out in the Comment \cite{Arrington:2015}, this modification 
is inconsistent with the analyticity constraints. The latter 
require that all the singularities of $G_E(Q^2)$ lie on the negative
$Q^2$--axis,  whereas the modification has a pole near the positive
axis resulting in a resonance-like structure, as seen in \Figref{GEminus1}
(red dashed curve), as well as in the figure of the Comment.

Here we present a modified $G_E$, shown in \Figref{GEminus1} (blue dotted curve), that complies with the
consistency requirements put forward in the Comment, and is yet
resolving the discrepancy in exactly the same way as described
in the original paper. 
The rest of this Reply can be viewed as the revised Sec.\ III of Ref.~\cite{Hagelstein:2015yma}:
\bigskip 

\centerline{\bf III. RESOLVING THE PUZZLE} 
\medskip 
\setcounter{equation}{19}

 We assume the electric FF to separate into a smooth ($\ol G_E$) and a nonsmooth part ($\widetilde G_E$), such that,
 \beq
  G_E(Q^2) = \ol G_E(Q^2) + \widetilde G_E(Q^2).\eqlab{newFF}
  \eeq 
 For the smooth part we shall take a well-known parametrization which fits the $ep$ data, while
 for the nonsmooth one we take
\beq
\widetilde G_E(Q^2)=\frac{A\,Q_0^2\,Q^2\left[Q^2+\eps^2\right]}{\left[Q_0^2+Q^2\right]^4},
\eeq
where $A$, $\eps$ and $Q_0$ are real parameters. The poles of this 
function are at negative $Q^2$ (timelike region) and hence it obeys the
analyticity constraint.

 \begin{figure}[tbh] 
    \centering 
       \includegraphics[scale=0.48]{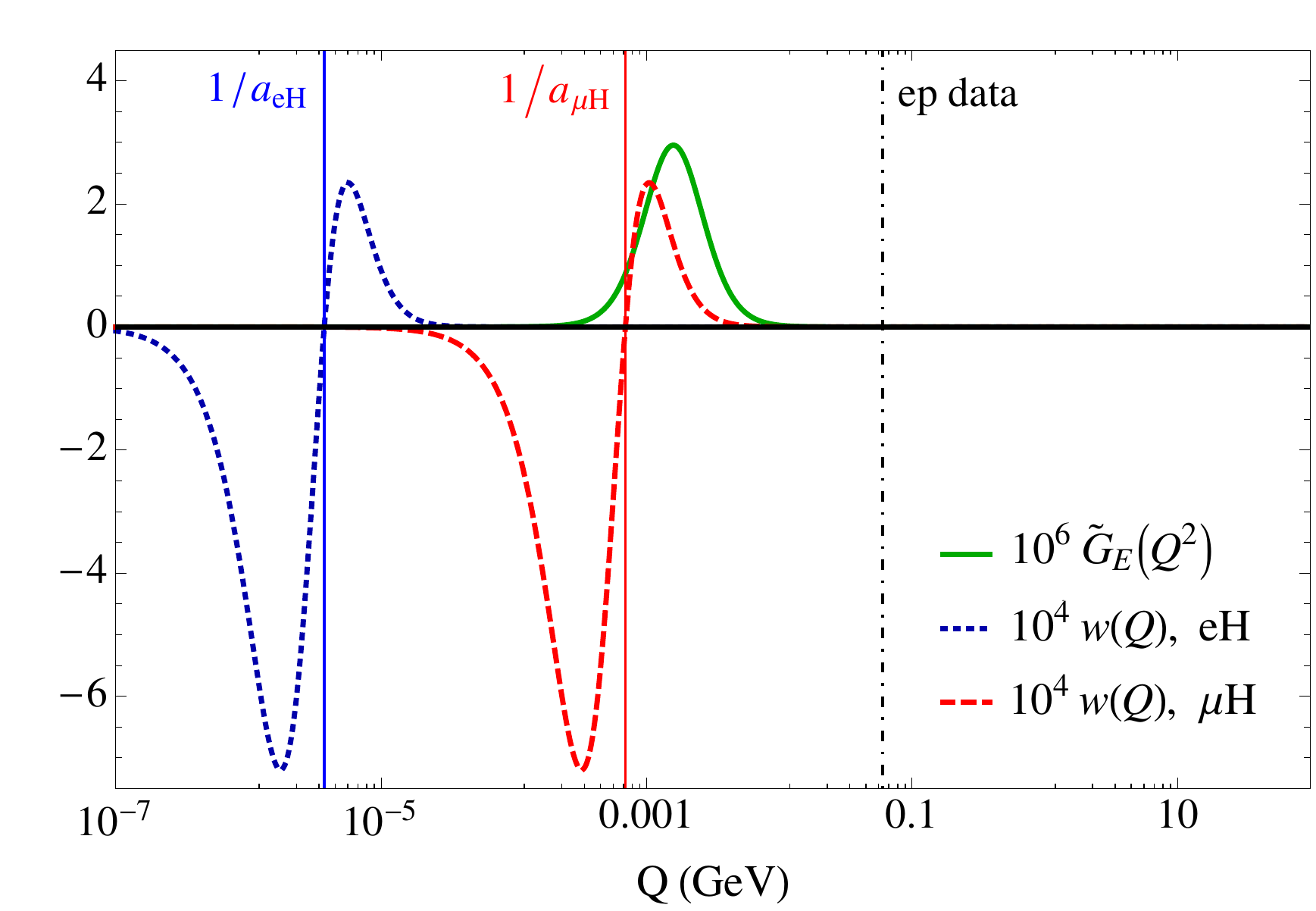}
       \caption{The correction, $\widetilde G_E(Q^2)$, for $Q_0=1.6$ MeV, $A=1.2\times10^{-4}$ MeV$^2$ and $\eps=0.143$ MeV (solid green), and the weighting function, $w(Q)$, for $e$H (blue dotted) and $\mu$H (red dashed) as functions of $Q$.  The dash-dotted line indicates the onset of electron-proton scattering data.}
              \label{fig:correction}
\end{figure}

According to \Figref{correction}, 
in order to make a maximal impact on the puzzle, the fluctuation $\widetilde G_E$ must be located at the extremi of $w(Q)$ in Eq.~(19a)\footnote{Equation numbers below 20 refer to the equations in
Ref.~\cite{Hagelstein:2015yma}.} around either the
$e$H or $\mu$H inverse Bohr radius. Here we shall only consider the latter case and set one of the position parameters to the MeV scale:
\beq
Q_0=1.6\, \mbox{MeV}.
\eeq
This choice conditions the choice of the smooth part $\bar G_E$, in case one wants to solve the puzzle.
Indeed, since with this $Q_0$ the nonsmooth part affects mostly the $\mu$H result, 
the smooth part must have a radius consistent with the $e$H value.
We therefore adopt the chain-fraction fit of Arrington and Sick \cite{Arrington:2006hm}:
\beq
\ol G_E(Q^2)=\frac{1}{1+\frac{3.478 \,Q^2}{1-\frac{0.140 \,Q^2}{1-\frac{1.311 \,Q^2}{1+\frac{1.128 \,Q^2}{1-0.233 \,Q^2}}}}} .
\eeq

Fixing $Q_0$, the other two parameters of $\widetilde G_E$,
 $A$, and $\eps$, are fitted by requiring our FF to yield the empirical 
Lamb shift contribution, in both normal and muonic hydrogen, i.e.:  
 \begin{subequations}
 \eqlab{LS}
 \bea
&& E^{\mathrm{FF}(exp.)}_{2P-2S}(e\mathrm{H}) = -0.620(11) \, \mbox{neV}, \label{LSeH}\\
&& E^{\mathrm{FF}(exp.)}_{2P-2S} (\mu\mathrm{H})  = -3650(2) \,  \mbox{$\upmu$eV} \label{LSmuH}.
\eea 
 \end{subequations}
Note that these are not the experimental Lamb shifts, but only the finite-size contributions, described by Eqs.~(2) and (4), with the corresponding empirical values for the radii. In the $e$H case we have taken the CODATA value of the proton radius, Eq.~(3a), which is obtained as an weighted average over several hydrogen spectroscopy measurements, and $R_{E(2)} = 2.78(14)$ fm \cite{Borie:2012zz}.  In the
 $\mu$H case we have taken the values from Ref.~\cite{Antognini:2013rsa}, hence
 Eq.~(3b) for the radius and the same as the above value for $R_{E(2)}$.

\begin{figure}[tbh] 
    \centering 
       \includegraphics[scale=0.48]{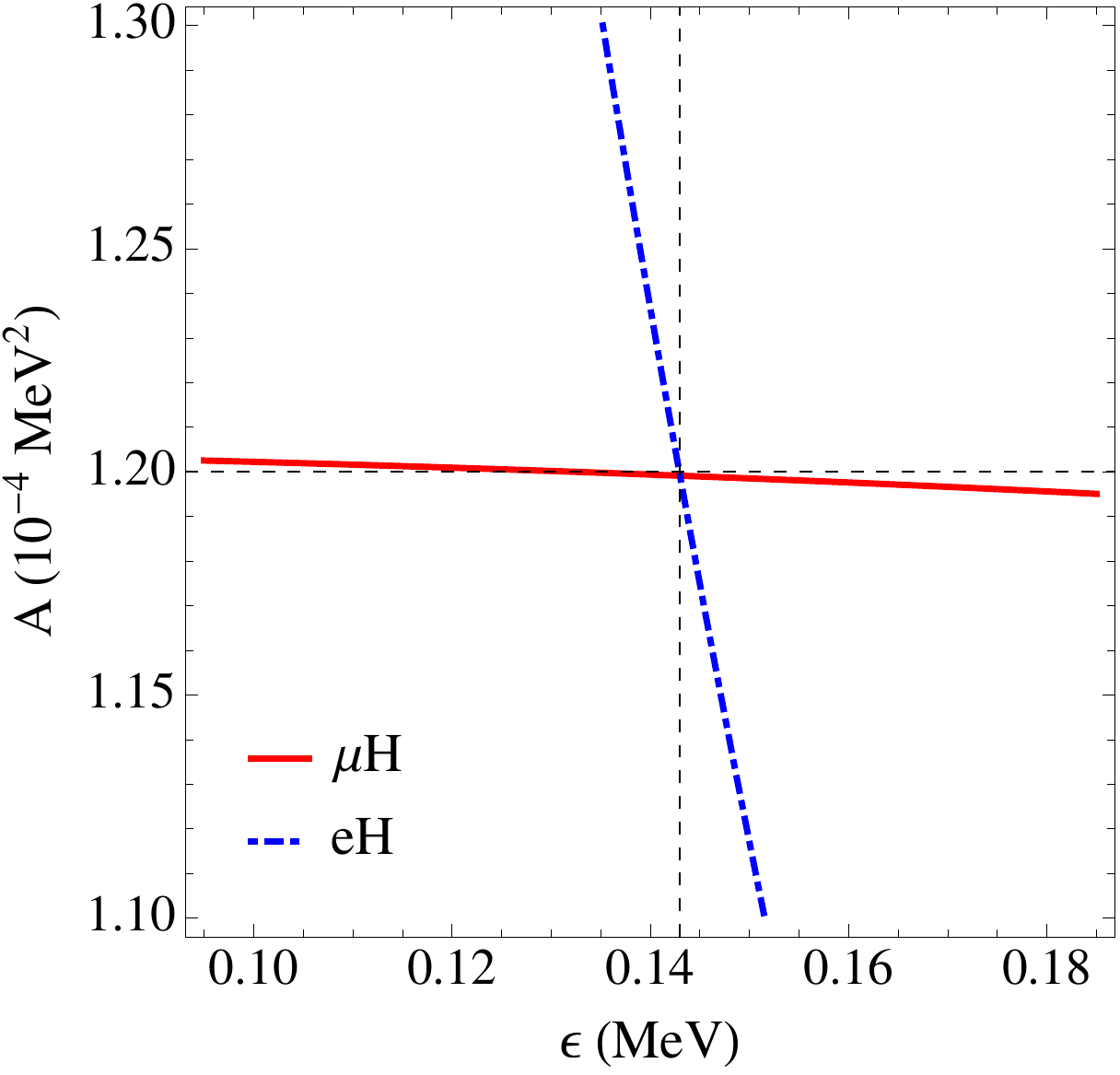}
       \caption{Parameters of $\widetilde G_E$ for which the $e$H and $\mu$H Lamb shifts of \Eqref{LS} 
       are reproduced. For fixed $Q_0=1.6$ MeV, we chose $A=1.2\times10^{-4}$ MeV$^2$ and $\eps=0.143$ MeV, as indicated by the dashed lines.}
              \label{fig:Parameter}
\end{figure}

Figure~\ref{fig:Parameter} shows at which $A$ and $\eps$ our
 FF complies with either the $e$H (blue dot-dashed curve) or $\mu$H (red solid curve) Lamb shift. For
 $A=1.2\times10^{-4}$ MeV$^2$ and $\eps=0.143$ MeV, our FF describes them both, thus resolving the puzzle (the description of the $ep$ data 
 by $\bar G_E$ is not affected by the addition of $\widetilde G_E$).

Figure \ref{fig:correction} shows the fitted $\widetilde G_E$, and the weighting function (17) for $e$H and $\mu$H. The modification thus enhances the FF in the region below the onset of $ep$ data ($Q<63$ MeV). The overlap between the correction and the positive contribution of the $\mu$H weighting function is clearly dominating, resulting in the desired matching to the experimental Lamb shifts given in \Eqref{LS}.

 We emphasize that the magnitude of the change in the FF is extremely tiny,
  \beq
\big\vert\widetilde G_E/\, \ol G_E\big\vert <3\times10^{-6},
\eeq
for any positive $Q^2$. The Comment suggests that a comparison of our correction to the deviation of the FF from unity is more fair. For our newly
 proposed $\widetilde G_E$, we find this ratio to be:
$$
\big\vert\widetilde G_E/\, (\ol G_E-1)\big\vert < 0.57\,,
$$
which does not seem unreasonable either. 
Furthermore, our new FF modification 
satisfies another criteria put forward in the Comment, namely: 
$G_E(Q^2) < 1$ for $Q^2 >0$.

Nevertheless, the modification obviously has a profound effect on the $\mu$H Lamb shift.
 Its effect on the second and third moments is given by:
\bea
\widetilde{\langle r^2\rangle}_E&\equiv &-6 \frac{\dd}{\dd Q^2} \widetilde G_E(Q^2)\Big|_{Q^2= 0}
=-\frac{6 A\eps^2}{Q_0^6},\quad\\
\widetilde{\langle r^3\rangle}_E&\equiv &\frac{48}{\pi} \int_0^\infty \!\frac{\dd Q}{Q^4}\,
\left\{ \widetilde G_E(Q^2) +\sixth \widetilde{\langle r^2\rangle}_E Q^2\right\},\nn\\
&=&15 A(Q_0^2-7\eps^2)/2Q_0^7.
\eea
The numerical values of these moments, together with their ``would be" effect on the Lamb shift and the non-expanded
Lamb result,  are given  
in Table \ref{Table}. One can see that the expansion in moments breaks down for the 
the modified FF contribution to $\mu$H.

 \begin{table}[tbh]
 \caption{Lamb shift and moments corresponding to our model FF, with $Q_0=1.6$ MeV, $A=1.2\times10^{-4}$ MeV$^2$ and $\eps=0.143$ MeV.}
 \label{Table}
\begin{tabular}{c|c|c|c|c}
&Eq.&$\ol G_E$&$\widetilde G_E$&$ G_E$\\
\hline
$\langle r^2\rangle_E \, [\mbox{fm}^2]$&(6a)&$(0.9014)^2$&$-(0.1849)^2$&$(0.8823)^2$\\
$\langle r^3\rangle_E \,[\mbox{fm}^3]$&(12)&$(1.052)^3$&$(8.539)^3$&$(8.544)^3$\\
\hline
Lamb-shift, expanded & (11) & && \\
$E_{2P-2S}^{\mathrm{FF}(1)}(e\mathrm{H})[\text{neV}]$ &&$-0.6569$&$0.0371$&$-0.6198$\\
$E_{2P-2S}^{\mathrm{FF}(1)}(\mu\mathrm{H})[\upmu\text{eV}]$&&$-4202$&$11542$&$7340$\\
\hline
Lamb-shift, exact & (19a) & && \\
$E_{2P-2S}^{\mathrm{FF}(1)}(e\mathrm{H}) [\text{neV}]$&&$-0.6569$&$0.0370 $&$-0.6200$\\
$E_{2P-2S}^{\mathrm{FF}(1)}(\mu\mathrm{H})[\upmu\text{eV}]$&&$-4202$&$552$&$-3650$\\
\end{tabular}
\end{table}

In conclusion, we have reworked the low-$Q$ modification of the empirical 
proton FF $G_E$ such that it complies with the criteria put forward in 
the Comment \cite{Arrington:2015}. The original (`old') and the reworked (`new')
modifications are shown in \Figref{GEminus1}, together with the
unmodified form. The old and new modification are quite different, yet they
both allow to describe the $e$H and $\mu$H Lamb shift simultaneously, while
maintaining the agreement with the $ep$ scattering data. The new modification
looks much more reasonable from the standpoint of the Comment. However, we emphasize 
once more that this is not a proposal for the solution of the puzzle --- not until 
a physical mechanism for this effect is found. For a current update on the 
status of the proton-radius puzzle, see \cite[Sec.\ 7]{Hagelstein:2015egb}
and references therein.

\section*{Acknowledgements}
This work was supported by the Deutsche Forschungsgemeinschaft (DFG) through the Collaborative Research Center SFB 1044 [The Low-Energy Frontier of the Standard Model], and the Graduate School DFG/GRK 1581
[Symmetry Breaking in Fundamental Interactions].


\begin{thebibliography}{99} 
\bibitem{Hagelstein:2015yma} 
  F.~Hagelstein and V.~Pascalutsa,
  {\it Breakdown of the expansion of finite-size corrections to the hydrogen Lamb shift in moments of charge distribution,}
  Phys.\ Rev.\ A {\bf 91}, 040502 (2015).
  
\bibitem{Arrington:2015} 
  J.~Arrington,
  {\it Comment on ``Breakdown of the expansion of finite-size corrections to the hydrogen Lamb shift in moments of charge distribution",}
  arXiv:1602.01461 [hep-ph] (to appear in Phys.\ Rev.\ A).



\bibitem{Arrington:2006hm}
J.~Arrington and I.~Sick,
{\it Precise determination of low-Q nucleon electromagnetic form factors and their impact on parity-violating e-p elastic scattering,}
Phys.\ Rev.\, C {\bf 76}, 035201 (2007).
  
  \bibitem{Borie:2012zz}
  E.~Borie,
  {\it Lamb shift in light muonic atoms: Revisited,}
  Annals Phys.\  {\bf 327}, 733 (2012).
 
  
  \bibitem{Antognini:2013rsa} 
  A.~Antognini, F.~Kottmann, F.~Biraben, P.~Indelicato, F.~Nez and R.~Pohl,
  {\it Theory of the 2S-2P Lamb shift and 2S hyperfine splitting in muonic hydrogen,}
  Annals Phys.\  {\bf 331}, 127 (2013).

\bibitem{Hagelstein:2015egb} 
  F.~Hagelstein, R.~Miskimen and V.~Pascalutsa,
  {\it Nucleon polarizabilities: From Compton scattering to hydrogen atom,}
  Prog.\ Part.\ Nucl.\ Phys.,
doi:10.1016/j.ppnp.2015.12.001  (arXiv:1512.03765 [nucl-th]).
\end{thebibliography}
\end{document}